\documentclass[aps,prl,reprint,superscriptaddress,twocolumn,showpacs,amsmath,amssymb, amsfonts,10pt]{revtex4-1}

\usepackage{amsmath}
\usepackage{amsfonts}
\usepackage{amssymb}
\usepackage{mwe}
\usepackage{graphicx}
\usepackage{color}
\usepackage{bm}
\usepackage{soul}

\usepackage[colorlinks,bookmarks=false,urlcolor=blue, citecolor=blue, linkcolor = blue]{hyperref}

\newcommand{\bb}[1]{\left( #1 \right)}
\newcommand{\mrm}[1]{\mathrm{#1}}

\newcommand{\be}{\begin{equation}}
\newcommand{\ee}{\end{equation}}

\definecolor{darkgreen}{rgb}{0.53, 0.66, 0.42}

\newcommand{\bra}[1]{\langle #1|}
\newcommand{\ket}[1]{|#1\rangle}
\newcommand*\diff{\mathop{}\!\mathrm{d}}

\usepackage[normalem]{ulem}

\begin{document}
\title{
Excitonic Tonks-Girardeau and charge-density wave phases in monolayer semiconductors 
}

\author{Rafa\l{} O\l{}dziejewski}
\email{rafal.oldziejewski@mpq.mpg.de}
\affiliation{Max-Planck-Institute of Quantum Optics, Hans-Kopfermann-Straße 1, D-85748 Garching, Germany}
\affiliation{Munich Center for Quantum Science and Technology (MCQST), Schellingstr. 4, D-80799 M{\"u}nchen, Germany}
\author{ Alessio Chiocchetta}
\affiliation{Institute for Theoretical Physics, University of Cologne, Z\"{u}lpicher Strasse 77, 50937 Cologne, Germany}
\author{ Johannes Kn\"{o}rzer}
\affiliation{Max-Planck-Institute of Quantum Optics, Hans-Kopfermann-Straße 1, D-85748 Garching, Germany}
\affiliation{Munich Center for Quantum Science and Technology (MCQST), Schellingstr. 4, D-80799 M{\"u}nchen, Germany}
\author{Richard Schmidt}
\affiliation{Max-Planck-Institute of Quantum Optics, Hans-Kopfermann-Straße 1, D-85748 Garching, Germany}
\affiliation{Munich Center for Quantum Science and Technology (MCQST), Schellingstr. 4, D-80799 M{\"u}nchen, Germany}
\date{March 2021}

\begin{abstract}
Excitons in two-dimensional semiconductors provide a novel platform for fundamental studies of many-body interactions.
In particular, dipolar interactions between spatially indirect excitons may give rise to strongly correlated phases of matter that  so far have been out of reach of experiments.
Here, we show that excitonic few-body systems in atomically thin transition-metal dichalcogenides undergo a crossover from a Tonks-Girardeau to a charge-density-wave regime.
To this end, we take into account realistic system parameters and predict the effective exciton-exciton interaction potential. We find that the pair correlation function contains key signatures of the many-body crossover already at small exciton numbers and show that photoluminescence spectra provide readily accessible experimental fingerprints of these strongly correlated quantum many-body states.
\end{abstract}

\date{\today}

\maketitle

Quantum systems often reveal their most striking features in low spatial dimension.
Prime examples include fractional excitations \cite{Steinberg2007}, fermionization \cite{Rigol2005} and non-trivial topology \cite{McGinley2018}. One-dimensional quantum systems offer an experimental playground to explore this rich physics and significant advances with quantum optical systems such as ultracold atoms \cite{Gross2017} and molecules \cite{Balakrishnan2016} have enabled the control and study of complex quantum matter in the laboratory. In particular, these platforms permit the study of dipolar quantum gases \cite{Lahaye2009} which are governed by long-range magnetic interactions.
A plethora of states of matter can be explored with dipolar gases, including classical and quantum glass phases \cite{Lechner2013}, novel superfluid \cite{Macia2014} and supersolid phases \cite{Boettcher2019,Chomaz2019,Tanzi2019}, spin liquids \cite{Yao2018}, and exotic few-body complexes \cite{Wunsch2011}. While ultracold atomic and molecular systems benefit from high tunability and advanced readout techniques, their experimental investigation is, however, so far restricted to comparatively low densities or small dipole moments.

Solid-state quantum systems provide an alternative setting for the realization of strongly interacting systems in reduced dimensions.
Notably, recent progress in material science and device fabrication has enabled the development of pristine low-dimensional semiconductors, which have become a versatile platform for the study of quantum many-body physics \cite{Tuan2019,Regan2019,Smolenski2020,zhou2020signatures}.
Particularly transition-metal dichalcogenides (TMDs) attract growing interest due to their unique optoelectronic properties \cite{Manzeli2017,Wang2018}.
Owing to an optical band gap and reduced screening, they enable efficient light-matter interfaces and host strongly bound excitonic quasiparticles.
Exciton physics in these atomically thin nanomaterials opens up new possibilities for experimental study of few- and many-particle phenomena \cite{sidlerfermi2017,Courtade2017,Mueller2018,Sigl2020,Hoegele2021,li2020dipolar,kremser2020discrete}.
In particular, spatially indirect  excitons feature increased lifetimes and significant electric dipole moments, making them ideally suited for the study of dipolar interactions in previously unexplored parameter regimes \cite{Fogler2014,Calman2018,Mak2019,Schneider2021}.

\begin{figure}[t!]
\begin{center}
		\includegraphics[width=0.48\textwidth]{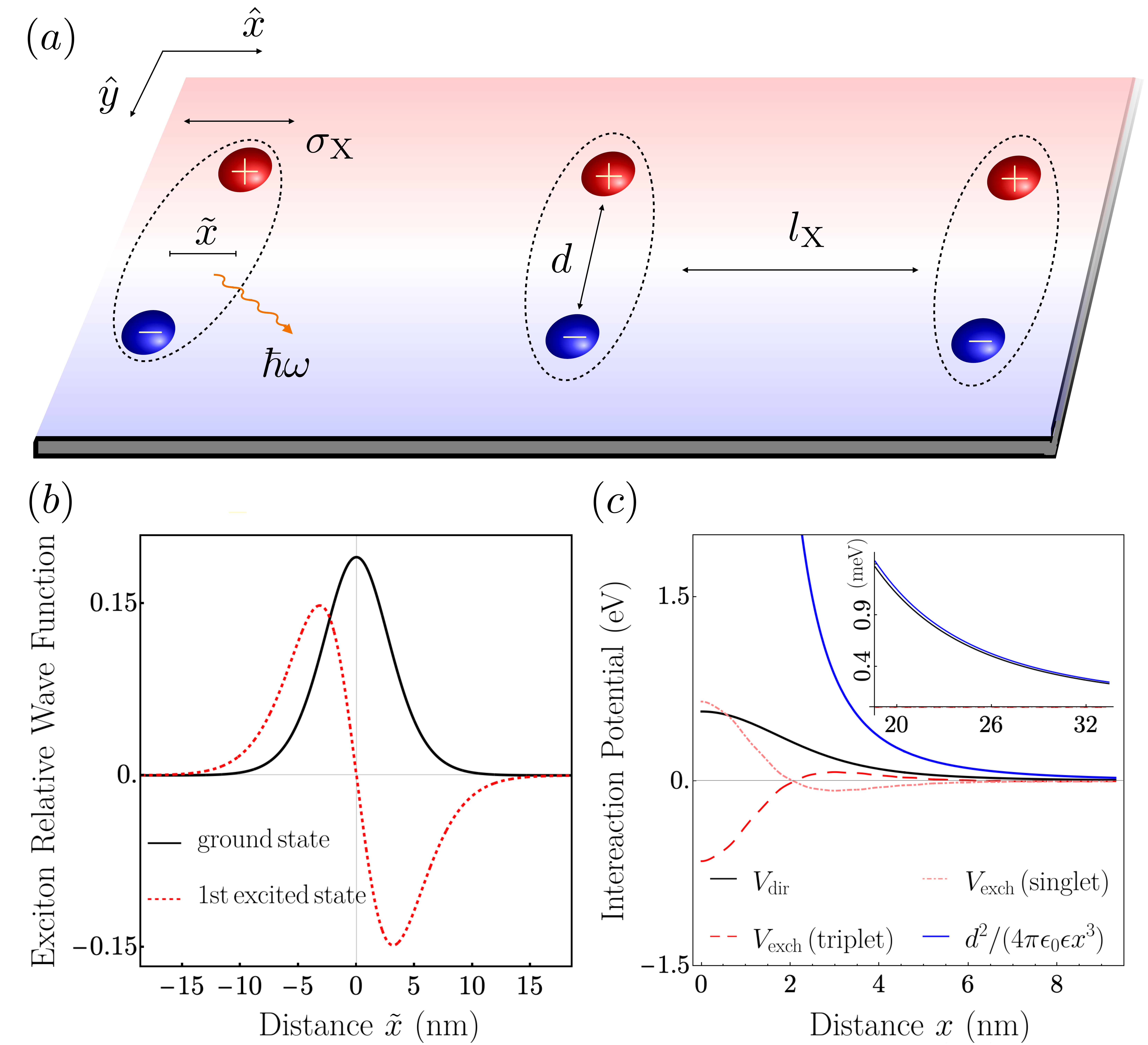}
\par\end{center}
\caption{(a) Sketch of the interfacial exciton system. Electrons and holes are separated by a distance $d \sim (1-4)$ nm.
Both the electron and the hole as well as the centre of mass of the resulting interfacial exciton  move along a straight line. The typical distance between the excitons $l_\mathrm{X}$ is much larger than $d$ and the size of a single exciton $\sigma_\mathrm{X}$.
(b) Relative motion wavefunction of the ground state (solid black line) and  1st excited state (red dotted line) of  the interfacial exciton.
(c) Contributions to the calculated interaction potential between two excitons.
At large distances, two excitons mutually interact via an effective potential $V_{\rm eff}(x)\approx \ (e d)^2/\bb{4 \pi \varepsilon_0\epsilon x^3}$, where $x$ is the inter-exciton separation.
\label{Fig:model}}
\end{figure}

In this Letter, we theoretically study the emergence of many-body physics in a one-dimensional few-exciton system with dipolar interactions.
We demonstrate that the system undergoes a crossover from a Tonks-Girardeau (TG)  to a charge-density wave (CDW) state even at moderate system sizes.
Owing to the advantageous material properties of TMDs, the  system can be operated in both regimes, depending on the exciton density.
Furthermore, we find that the photoluminesence (PL) spectrum of the system contains an optical fingerprint of this crossover that can be readily probed in state-of-the-art experiments.

\textbf{\textit{Setup and model.---}} 
We consider a one-dimensional system of $N$ dipolar excitons, as can be hosted at the interface of lateral heterojunctions in TMDs \cite{Lau2018}, with electrons and holes separated by a distance $d\sim(1-4)~\mathrm{nm}$ that may be tuned by a gate voltage, see Fig.~\ref{Fig:model}(a).
Further, we assume that the typical distance between the center-of-mass (COM) coordinates of two adjacent excitons $l_\mathrm{X}$ is large, i.e.~$l_\mathrm{X} \gg d$, and that the exciton wavefunction is highly localized in the $\hat y$ direction. Under these assumptions the exciton motion is constrained to a one-dimensional wire along $\hat x$.

\textbf{\textit{Single exciton.---}} 
The system of an electron and a hole is described by the Hamiltonian $H_{X} = -\hbar^2\partial^2_{x_e}/(2m_e)-\hbar^2\partial^2_{x_h}/(2m_h)+V_\text{K}(x_{eh})$,
with $m_{e,h}$ the effective electron and hole mass, respectively, and $x_{e,h}$ the electron and hole coordinates along the wire of length $L$. While we assume periodic boundary conditions for simplicity, our key results also apply to fixed boundary conditions. The Rytova-Keldysh potential $V_\text{K}$ models the dielectric screening of the Coulomb interaction~\cite{Cudazzo2011,VanTuan2018,suppmat}, and $x_{eh}\equiv \sqrt{(x_e-x_h)^2+d^2}$ is the distance between electron and hole. Owing to their transversal separation, the electron and hole can be assumed to be distinguishable particles, i.e., exchange effects are neglected, as well as valence-band mixing due to spin-orbit interaction, dynamical screening, the finite width of the TMD monolayer and the transversal motion of electrons and holes along the $\hat y-$direction.

The eigenstates of $H_X$ have the form $\Phi^{Q}_{n}(x_{e},x_h)=\phi_{n}(\tilde x)\mrm{e}^{iQX}/\sqrt{L}$, with $\tilde x=x_e-x_h$ the relative coordinate, $X$ the center of mass and $Q$ the total momentum. The wave function for the internal, relative motion of the interfacial exciton $\phi_{n}(\tilde x)$ is obtained from $[-\hbar^2\partial^2_{\tilde x}/(2\mu) + V_\text{K}(x_{eh})] \phi_{n}=E_{n} \phi_{n}$, where $\mu = m_em_h/(m_e+m_h)$ denotes the reduced mass. Fig.~\ref{Fig:model}(b) shows $\phi_{n}(\tilde x)$ for the ground and first excited states, obtained from exact diagonalization using representative material parameters for MoSe$_2$  monolayers~\cite{kylanpaa2015binding,Fey2020}, $d=4$~nm and an average dielectric constant $\epsilon=2.5$.
Both states are bound.
The ground state (GS) energy is approximately $110$ meV and the gap to the first excited state is $\sim30$ meV.
The width of the ground-state wavefunction is $\sigma_\mathrm{X}  \approx 4.5$ nm~\footnote{We define $\sigma_\mathrm{X}$ as full width at half maximum of $|\phi_{\rm{GS}}(\tilde x)|^2$}.

\textbf{\textit{Two excitons.---}}
Due to their large binding energy and  long lifetime,  interfacial excitons as described above are ideal building blocks for studying the many-body physics of interacting bosons in TMDs. A key ingredient for their description is the derivation of the inter-exciton interaction potential, which we model  using an approach developed in the context of bulk semiconductors~\cite{ciuti1998role,tassone1999exciton,inoue2000renormalized,de2001exciton,okumura2001boson,schindler2008analysis,meyertholen2008biexcitons}, and recently applied to monolayer TMDs~\cite{shahnazaryan2017exciton}; for details see \cite{suppmat}.
At low carrier density and temperature the system is governed  predominantly by low-energy scattering and the mutual interactions $\hat{V}_{\mrm{int}}$ between the fermionic constituents of the excitons can be treated as a perturbation to a system of two non-interacting excitons.
Moreover, at sufficiently large inter-exciton separation, one may assume that only the GS exciton is occupied owing to the large gap to the first excited state.
Under these conditions, the COM momenta of excitons are small and, consequently, we consider the scattering of two ground-state excitons  focussing only on leading terms in the effective potential as given by the Hartree-Fock and Born approximations.
Assuming that excitons have the same parallel spin projections, the scattering process schematically reads $
\bb{GS,Q}+\bb{GS,Q'}\xrightarrow{V_{\mrm{eff}}(q)} \bb{GS,Q+q}+\bb{GS,Q'-q}$~\cite{suppmat}.
 Here, $V_{\mrm{eff}}(q)$ is the sought-after effective potential, and $q$ denotes the exchange momentum between the two excitons.
The effective potential $V_{\mrm{eff}}(q)=V_{\mrm{dir}}(q)+V_{\mrm{exch}}(q)$ consists of a direct part $V_{\mrm{dir}}$ that relates to the classical interaction between two electric dipoles and the bosonic exchange of the excitons, while $V_{\mrm{exch}}$ accounts for fermionic exchanges between the two holes and two electrons.

In Fig.~\ref{Fig:model}(c), we present the contributions to the total exciton-exciton interaction, $V_{\mrm{dir}}(x)$ and $V_{\mrm{exch}}(x)$, for different spin states of fermions. At large distances, the direct part of the interaction  reduces to the classical dipolar potential $V_{\mrm{dir}}(x) \xrightarrow{x\rightarrow\infty} A (e d)^2/\bb{4 \pi \varepsilon_0\epsilon x^3}$ (see the inset in Fig.~\ref{Fig:model}(c)), slightly reduced in strength by a factor $A<1$ that results from both corrections to the Coulomb interactions encoded in the Keldysh potential and the finite extent of the interfacial exciton wave function. 
Due to the interfacial character of the excitons, the exchange part contributes to the effective potential only at short distances and decays exponentially. For $d\approx (1-4)$ nm, we find that the  range of the exchange potential $r_{\mrm{exch}}$ is of the same order as the size of the ground state, $r_{\mrm{exch}}\sim\sigma_{X}$. It thus can be neglected, i.e. $V_{\mrm{eff}}(x)=V_{\mrm{dir}}(x)$ ($l_\mathrm{X}=n^{-1}=L/N \gg \sigma_\mathrm{X}$).
Crucially, one can check {\it a posteriori} that  the  interaction energy per particle in the many-body system does not exceed a few meV, thus confirming the validity of our assumptions.

\textbf{\textit{Many excitons.---}} 
So far, we have argued that ground state interfacial excitons can be treated as rigid bosons with mass $m_X=m_e+m_h$, whose internal structure is robust against mutual interactions. Now, we shall investigate $N$ such particles moving in a periodic wire of length $L$. This effective 1D system is governed by the Hamiltonian
\be
H=-\frac{\hbar^2}{2m_{X}}\sum_{i=1}^{N} \frac{\partial^2}{\partial x_i^2}+\sum_{i<j}^{N} V_{\mrm {eff}}(x_i-x_j),\label{eq:hamgen}
\ee
with $x_i$ denoting the exciton coordinates.

\begin{figure}[t!]\label{Fig:phase_diagram}
\begin{center}
		\includegraphics[width=0.49\textwidth]{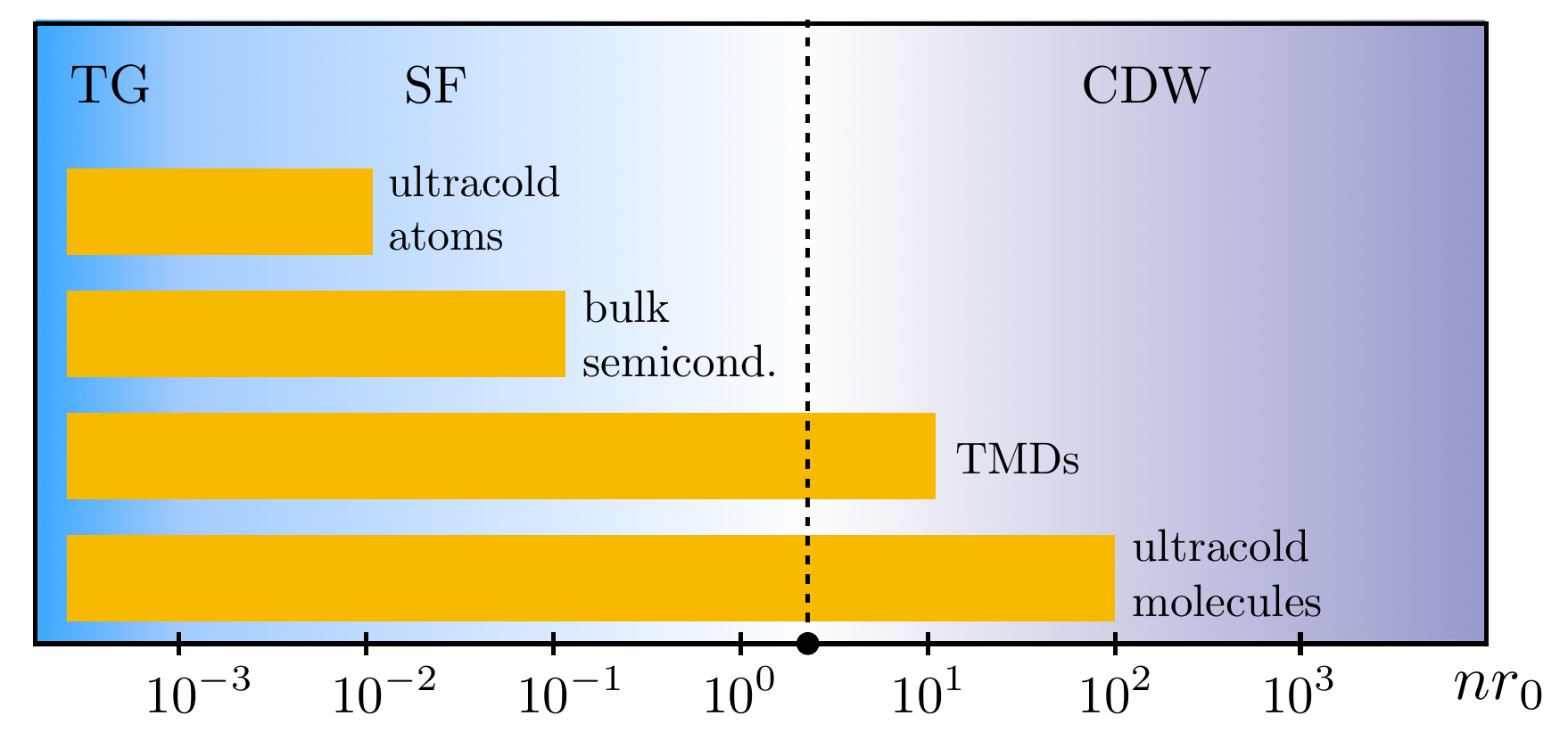}
\par\end{center}
\caption{ Illustration of the zero-temperature phase diagram of a one-dimensional dipolar quantum gas and a comparison of its accessibility using different experimental platforms, based  on~\cite{Arkhipov2005,Citro2007,Citro_2008,Dalmonte2010} and the present study of TMDs. TMDs can be expected to outperform ultracold atomic gases (e.g., Cr, Er and Dy) and bulk semiconductors (e.g., GaAs) and complement ultracold molecules.
\label{Fig:phase-diagram}}
\end{figure}

At low temperatures, the properties of the exciton wire are encoded in the ground state of Eq.~\eqref{eq:hamgen}.
Given the dipolar character of $V_{\mrm {eff}}(x)$ at intermediate and large distances, 
insight into 
the ground state can be gained from the study of ultracold dipolar gases~\cite{Arkhipov2005,Citro2007,Citro_2008,Dalmonte2010}. The phase diagram is controlled by the single dimensionless parameter $nr_0$ where the dipolar length $r_0=m_X(ed)^2/\bb{2\pi\hbar^2\varepsilon_0\epsilon}$ characterizes the dipolar interaction strength~\footnote{Note that, in general, $V_{\mrm{dipolar}}(x)=\frac{C_{\mrm{dd}}}{4\pi}\frac{1}{x^3}$, and thus $r_0=\frac{mC_{\mrm{dd}}}{2\pi\hbar^2}$.
The interaction strength depends on the origin of the dipolar interaction (magnetic vs electric dipoles)} and  $n=N/L$ the 1D density of excitons.
In the thermodynamic limit, the system undergoes a crossover from a Tonks-Girardeau (TG) state at low density ($nr_0\rightarrow 0$) via a strongly correlated superfluid (SF) at intermediate $nr_0$ to a charge-density-wave (CDW) state at large densities, see Fig.~\ref{Fig:phase-diagram}. In TMDs all regimes are accessible: for $d\sim (1-4)$~nm and $\epsilon \sim 2.5 $, we find $r_0 = (20 - 320)$~nm.
In order to satisfy the dipolar gas approximation, exciton densities in the system have to fulfill $n^{-1}\gg \sigma_\mathrm{X}$, implying $n \lesssim 10^6 \, \mrm{cm}^{-1}$, corresponding to a two-dimensional density $n_{\mrm{2D}}\lesssim 10^{12} \, \mrm{cm}^{-2}$, consistent with typical exciton densities in TMDs~\cite{tan2020interacting}.
Therefore, experiments can in principle reach up to $nr_0 \lesssim 32$.
As illustrated in Fig.~\ref{Fig:phase-diagram}, TMDs can thus outperform most alternative experimental platforms and complement studies using ultracold molecules that have yet to reach the required densities and control in order to realize their full potential.

Using currently available nano-fabrication techniques,  excitons will be trapped in finite size systems so that only systems containing  a finite number of excitons are accessible experimentally~\cite{li2020dipolar,kremser2020discrete,Forg2021}. In the following, we  show that the TG-CDW crossover leaves its fingerprints in excitonic wires even in such finite, few-body systems, where the exciton density can be controlled by laser intensity. In order to theoretically study this scenario, we evaluate the many-body eigenstates of the Hamiltonian \eqref{eq:hamgen} using
exact diagonalization~\cite{suppmat}.

\begin{figure}[t!]
\begin{center}
		\includegraphics[width=0.49\textwidth]{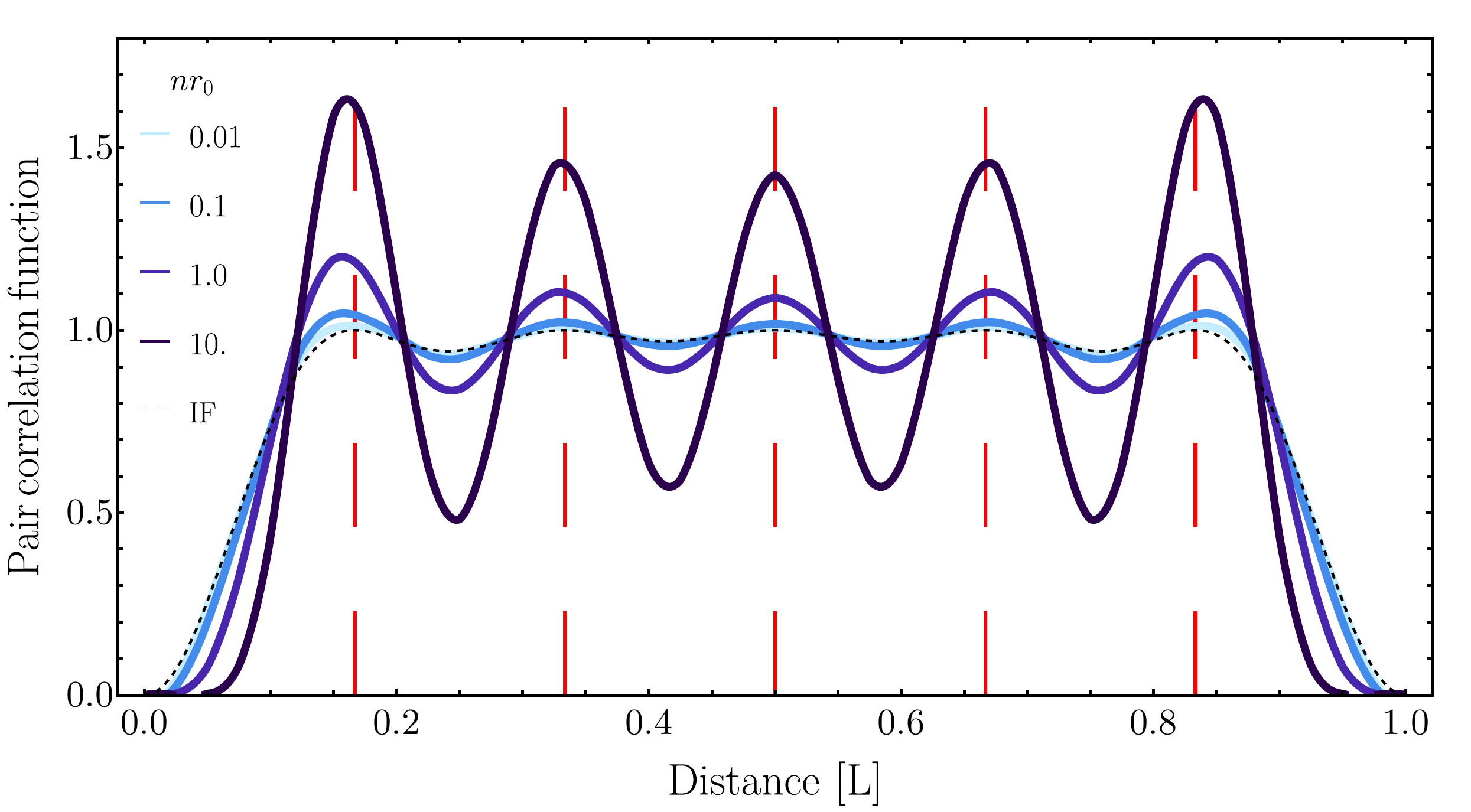}
\par\end{center}
\caption{Pair correlation function as a function of distance at fixed exciton number  $N=6$ for different $nr_0$ tuned by adjusting the system size $L$.
The result for ideal fermions (IF) is recovered in the TG limit for $nr_0\to 0$.
\label{Fig:g2}}
\end{figure}
One quantity that signals the TG-CDW crossover \cite{Arkhipov2005,knap2012clusterd} is provided by the pair correlation function
\begin{equation}
    g^{(2)}(x) = \frac{\langle \psi^\dagger(0)\psi^\dagger(x)\psi(x)\psi(0)\rangle}{\langle \psi^\dagger(x)\psi(x) \rangle\langle \psi^\dagger(0)\psi(0) \rangle},
\end{equation}
with $\psi(x)$ and $\psi^\dagger(x)$ the exciton annihilation and creation operators in real space.
In Fig.~\ref{Fig:g2} we show $g^{(2)}(x)$  for a fixed number of $N=6$ excitons for different values of $nr_0$ by tuning the size of the system $L$.
\begin{figure*}[ht!]
\begin{center}
		\includegraphics[width=0.98\textwidth]{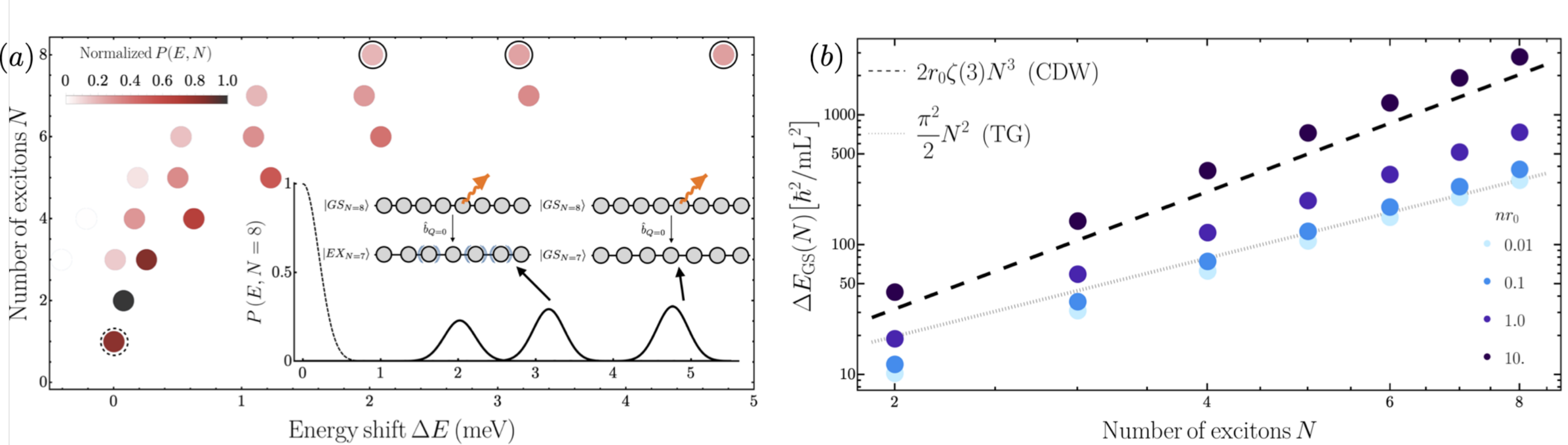}
\par\end{center}
\caption{(a) Photoluminescence (PL) spectrum  as function of detuning  from the bare exciton resonance at fixed system size $L=156$ nm and dipolar length $r_0=280$ nm that for $N=6$ yields $nr_0=10$.
Only the most dominant PL peaks are shown, the strongest corresponding to a ground-ground state transition $| GS_N\rangle \rightarrow| GS_{N-1}\rangle$, while  the rest originates from exciton decays leading to excited states of the systems. Inset: The PL spectrum for $N=8$ (solid line). The bare exciton resonance is shown as a dashed line. (b) Blue shift $\Delta E_{\mrm{GS}} (N)=E_{\mrm{GS}}(N)-E_{\mrm{GS}}(N-1)$ of the ground-to-ground state transition for $r_0=280$ nm as a function of $N$ for different values of $nr_0$ (values in the legend calculated for $N=6$). 
\label{Fig:plspectra}}
\end{figure*}
Crucially, even at such low particle numbers the observed features are in qualitative agreement with those of dipolar ultracold bosons 
in extended systems~\cite{Arkhipov2005,knap2012clusterd}.   
For $x \to 0$, $g^{(2)}(x)$ vanishes independently of the value of $nr_0$, as a consequence of the strong repulsive interaction at short range.
For small values of $nr_0$, the function $g^{(2)}(x)$ approaches unity for increasing $x$, with small superimposed Friedel-type oscillations.
The function is nearly indistinguishable from that for free fermions, reflecting the fact that excitons are fermionized in the TG regime and exhibit a liquid-like character. For larger $nr_0$, the oscillations increase in magnitude, signalling the emerging crystalline CDW phase.
The positions of the peaks almost coincide with the lattice sites in the solid phase (corresponding to 
$n r_0 \rightarrow \infty$), shown as  dashed vertical lines in Fig.~\ref{Fig:g2}. 

While this result shows that one can observe the crossover in a few-body system ($N=6$), $g^{(2)}(x)$ is challenging to access with current experimental techniques. Thus the question arises which other indirect observable might provide a probe of the TG to CDW crossover.
We hence shift our focus to another distinct signature of the phases in a 1D dipolar gas.
A straightforward, analytical  analysis for a large system shows  that the GS energy exhibits two limiting cases.
As shown in~\cite{suppmat}, for the TG state the GS energy  shows a  scaling with exciton number, $E_{\mrm{GS}}(N)\sim N^3$, that is distinct from the CDW state where $E_{\mrm{GS}}(N)\sim N^4$ is found.
As we will demonstrate in the following, these different scalings manifest themselves already at small exciton numbers in PL spectra that are readily accessible in experiments.

\textbf{\textit{The photoluminescence spectrum.---}} In TMDs, optical emission is one of the prime exciton decay mechanisms when  exciton densities are sufficiently low to avoid Auger processes. The probability of emission of a PL photon is the highest for  excitons with center-of-mass momentum inside the light-cone, i.e. $Q\sim 0$~\cite{Savona1999}. Since the typical equilibration time of the system is much shorter than the lifetime of an interfacial exciton this allows us to model the PL spectra within Fermi's Golden Rule as
\begin{equation}
\label{eq:PL_definition}
    P(E,N) \sim \sum_{k} |\langle k_{N-1} | \hat{b}_{Q=0} | GS_N\rangle|^2 \delta (E - E_{\mrm{GS}}(N) + E_k ),
\end{equation}
with $\hat b_Q$ the exciton annihilation operator at momentum $Q$,  $E$ the energy detuned from the bare exciton resonance, $| GS_N\rangle$ the GS of the system of energy $E_{\mrm{GS}}(N)$ containing $N$ excitons, $|k_{N-1}\rangle$ the eigenstates (including the GS) of the system with $N-1$ excitons, and $E_k$ the corresponding energies. 

In Fig.~\ref{Fig:plspectra}(a) we present the dominant peaks in the PL spectra upon varying the total exciton number $N$, for fixed system size $L$.
The system size is chosen such that for $N=6$ one obtains a dimensionless interaction parameter $nr_0=10$~\footnote{Note that for each $N$, we present here only the three most visible peaks.
In all cases considered in this work, the fourth-strongest peak would be invisible to the eye.}. In the inset  we show as an example an individual PL spectrum $P(E,N)$ for $N=8$. The observed structure holds for all densities: we observe a blue shift due to the repulsive interaction, and the most probable transition corresponds to $| GS_N \rangle \rightarrow | GS_{N-1} \rangle$, while the other peaks are related to transitions to low-lying excitations of the system. We observe that the blue shift given by $\Delta E_{\mrm{GS}} (N)=E_{\mrm{GS}}(N)-E_{\mrm{GS}}(N-1)$ increases for larger densities with a non-linear dependence.

In Fig.~\ref{Fig:plspectra}(b), we inspect the blue shift more closely by comparing our few-body results for different $nr_0$ with the limiting cases for a large system, where $\Delta E^{\mrm{TG}}_{\mrm{GS}}\sim N^2$ and $\Delta E^{\mrm{CDW}}_{\mrm{GS}}\sim N^3$~\cite{suppmat}. Remarkably, we observe not only a stark difference between the scaling of $\Delta E_{\mrm{GS}}$ for varying density but, indeed, recover the scaling behaviour of the many-body system.
Note that the offset of the $N^3$ power law can be attributed to the finite spread of the $g^{(2)}(x)$ function. The scaling behavior of the PL spectra shows that already surprisingly small exciton systems are sufficient to reveal fingerprints of the underlying correlated phases governing the physics in the large-system limit. 

\textbf{\textit{Conclusions and outlook.---}}
 We have demonstrated that interfacial excitons provide a promising platform for studying strongly correlated many-body physics.
 In particular, confinement in one-dimensional geometries gives rise to a dipolar Bose gas at interaction strengths that are out of reach of ultracold atomic or bulk semiconductor realizations.
We have shown that, given realistic experimental parameters, the full crossover between a fermionized Tonks-Girardeau phase and a charge-density-wave phase can be realized.
We demonstrated that signatures of these correlated states are measurable in conventional PL experiments.
Remarkably, already systems containing a small number of excitons ($N$ less than ten) are sufficient to identify clear fingerprints of this crossover between many-body states of matter that so far has eluded experimental observation.

Our work opens novel avenues for solid-state based quantum simulation: for example, lattice effects can be implemented by superimposing Moir\'{e} patterns and interactions may be tunable using Feshbach resonances \cite{FeshbachPolariton,kuhlenkamp2021,Wagner2021}.
Moreover, the coupling between excitons and the photonic modes of a cavity can give rise to exciton-polaritons, whose lighter mass may enable higher degeneracy temperatures.
An important question arises whether also in such systems one can study signatures of many-body effects using an ensemble of only few exciton polaritons. Finally, coupling quantum emitters (QE)~\cite{Srivastava2015,He2015,Palacios2017} to  excitonic wires represents an intriguing possibility to establish quantum coherence between different QEs, opening perspectives to engineer many-body superradiant phases and enable novel functionalities for quantum photonics technologies.  
\begin{acknowledgments}
\textbf{\textit{Acknowledgments.---}} We thank  J. Finley, A. Imamo\u{g}lu,  A. Srivastava and A. Stier  for stimulating discussions. R.~O. also thanks V. C. Arizona for fruitful discussions. 
R.~O., J.~K., and R.~S. are supported  by the Deutsche Forschungsgemeinschaft (DFG, German Research Foundation) under Germany’s Excellence Strategy – EXC-2111 – 390814868. A.~C. acknowledges support by the funding from the European Research Council (ERC) under the Horizon 2020 research and innovation programme, grant agreement No. 647434 (DOQS).

\end{acknowledgments}

\bibliography{biblio}

\clearpage

\begin{widetext}
\appendix
\section*{SUPPLEMENTAL MATERIAL}

\section{I. Derivation of the exciton-exciton interaction potential $V_{\mrm{eff}}(r)$}
\label{A:interaction}
In order to derive the exciton-exciton interaction potential for the interfacial system we use methods successfully employed for other solid-state systems, for example in quantum wells or for monolayer TMDs~\cite{ciuti1998role,tassone1999exciton,inoue2000renormalized,de2001exciton,okumura2001boson,schindler2008analysis,meyertholen2008biexcitons,shahnazaryan2017exciton}. We shall briefly summarize our approach following the reasoning presented in Ref.~\cite{shahnazaryan2017exciton}. 

In general, the elastic scattering process between two 1D interfacial excitons with momenta $Q$ and $Q'$ in the same internal state $n$ and parallel spin projections, involving the transfer of the wave vector $q$ can be schematically written as
\be\label{scattering}
\bb{n,Q}+\bb{n,Q'}\rightarrow \bb{n,Q+q}+\bb{n,Q'-q}.
\ee
The two-exciton wave function has to be antisymmetric under exchanges of the two electrons and two holes separately.
This directly  ensures that the total wave function is symmetric under the simultaneous exchange of the two electrons and two holes, as it should be for the composite bosons.
Here, we neglect electron-hole exchange effects so that electrons and holes are treated as distinguishable quasi-particles. Moreover, we do not include valence-band mixing due to spin-orbit interaction, assuming a two-band model within the effective mass approximation. Note that since electrons and holes have both spin and spatial degrees of freedom, the condition for the total wave function being antisymmetric under the exchange of alike charge carriers can only be satisfied when their spin and spatial parts have opposite symmetries.
Taking all the above into account, the two-exciton initial state of the scattering process \eqref{scattering} within the Hartree-Fock approximation reads
\begin{align}\label{twoexwave}
\Psi^{Q,Q'}_{n}(x_e,x_h,x_{e'},x_{h'})
&= \frac{1}{2\sqrt{2}}\left[ \Phi^{Q}_{n}(x_e,x_h) \Phi^{Q'}_{n}(x_{e'},x_{h'}) + \Phi^{Q}_{n}(x_{e'}, x_{h'})\Phi^{Q'}_{n}(x_e,x_h) \right] \notag \\
&\mp \frac{1}{2\sqrt{2}}\left[ \Phi^{Q}_{n}(x_{e'},x_h)\Phi^{Q'}_{n}(x_e,x_{h'})+ \Phi^{Q}_{n}(x_e,x_{h'})\Phi^{Q'}_{n}(x_{e'},x_h) \right],
\end{align}
where the $\mp$ sign refers to both excitons in the singlet states, for which both electrons and holes might be in a triplet (minus sign) or singlet (plus sign) spin states. For the excitons being in the triplet state, only the minus sign applies.

The scattering amplitude $V_{n}(Q,Q',q)$ characterizing a scattering event in Eq.~\eqref{scattering} is expressed as
\be\label{scatamplitude}
V_{n}(Q,Q',q)=\bra{\Psi^{Q,Q'}_{n}}\hat{V}_{\mrm{int}} \ket{\Psi^{Q+q,Q'-q}_{n}},
\ee
which is given by the following expression:
\begin{align}
V_{n}(Q,Q',q) &= \int \diff x_e \diff x_h \diff x_{e'} \diff x_{h'} {\Psi_{n}^{Q,Q'}}^*(x_e,x_h,x_{e'},x_{h'}) V_{\mrm{int}}(x_e,x_h,x_{e'},x_{h'}) \Psi^{Q+q,Q'-q}_{n}(x_e,x_h,x_{e'},x_{h'}) \notag \\
&= \frac{1}{2}\int \diff x_e \diff x_h \diff x_{e'} \diff x_{h'}{\Phi^{Q}_{n}}^*(x_e,x_h){\Phi^{Q'}_{n}}^*(x_{e'},x_{h'}) V_{\mrm{int}}(x_e,x_h,x_{e'},x_{h'}) \Phi^{Q+q}_{n}(x_e,x_h)\Phi^{Q'-q}_{n}(x_{e'},x_{h'}) \notag \\
&+ \frac{1}{2}\int \diff x_e \diff x_h \diff x_{e'} \diff x_{h'}{\Phi^{Q}_{n}}^*(x_e,x_h){\Phi^{Q'}_{n}}^*(x_{e'},x_{h'}) V_{\mrm{int}}(x_e,x_h,x_{e'},x_{h'}) \Phi^{Q+q}_{n}(x_{e'},x_{h'})\Phi^{Q'-q}_{n}(x_e,x_h) \notag \\
&-\frac{1}{2} \int \diff x_e \diff x_h \diff x_{e'} \diff x_{h'}{\Phi^{Q}_{n}}^*(x_e,x_h){\Phi^{Q'}_{n}}^*(x_{e'},x_{h'}) V_{\mrm{int}}(x_e,x_h,x_{e'},x_{h'}) \Phi^{Q+q}_{n}(x_{e'},x_h)\Phi^{Q'-q}_{n}(x_e,x_{h'}) \notag \\
&-\frac{1}{2} \int \diff x_e \diff x_h \diff x_{e'} \diff x_{h'}{\Phi^{Q}_{n}}^*(x_e,x_h){\Phi^{Q'}_{n}}^*(x_{e'},x_{h'}) V_{\mrm{int}}(x_e,x_h,x_{e'},x_{h'}) \Phi^{Q+q}_{n}(x_e,x_{h'})\Phi^{Q'-q}_{n}(x_{e'},x_h). 
\end{align}
Within the Hartree-Fock and Born approximations, the scattering amplitude \eqref{scatamplitude} represents the sought-after effective potential describing the interaction between two interfacial excitons for a specific scattering channel, e.g. for a given $n$.
We can immediately rewrite Eq.~\eqref{scatamplitude} in a more instructive way as
\begin{equation}\label{scatamplitude1}
V_{n}(Q,Q',q) = V_{n}^{\mrm{dir}}(Q,Q',q)+V_{n}^{\mrm{XX}}(Q,Q',q)+ V_{n}^{\mrm{ee}}(Q,Q',q)+V_{n}^{\mrm{hh}}(Q,Q',q),
\end{equation}
where superscripts XX, ee, and hh denote the exciton, electron, and hole exchange interaction parts, respectively, of the total interaction for a given scattering channel. So far we assumed nothing about the specific form of interaction between particles, and the above approach applies, in principle, to any two-body potential.
In particular, the generalization of the method summarized and employed here is straightforward for two excitons in different internal states as well as for scattering processes involving spin flips of the charge carriers.
In the most general case one has to diagonalize the full scattering matrix with the matrix elements given by Eq.~\eqref{scatamplitude}.

Typically,  interaction potentials in condensed matter are conservative. This also applies to our case so that we have:
\begin{equation}
    V_{\mrm{int}}(x_e,x_h,x_{e'},x_{h'})=V_{\mrm{int}}(x_e-x_{e'},x_h-x_{h'},x_{e}-x_{h'},x_{h}-x_{e'
    }).
\end{equation}
Therefore, using Eq.~\eqref{twoexwave}, we can simplify the direct part of interaction $V_{n}^{\mrm{dir}}(Q,Q',q)$ according to
\begin{align}\label{intdirect}
   & V_{n}^{\mrm{dir}}(Q,Q',q)=V_{n}^{\mrm{dir}}(q)= \\ \notag & \frac{1}{2L}\int \diff \tilde x \diff \tilde x' \diff \eta \, \mrm{e}^{i q \eta}|{\phi_{n}(\tilde x)}|^2|{\phi_{n}(\tilde x')}|^2 V_{\mrm{int}}(|\eta+\beta_h (\tilde x- \tilde x')|,|\eta-\beta_e (\tilde x-\tilde x')|,|\eta+\beta_h \tilde x +\beta_e \tilde x'|,|\eta-\beta_h \tilde x -\beta_e \tilde x'|),
\end{align}
where we have introduced the notations
$\tilde x=x_e-x_h$, $X=\beta_ex_e+\beta_hx_h$, $\tilde x'=x_{e'}-x_{h'}$, $X'=\beta_e x_{e'}+\beta_h x_{h'}$, $\eta=X-X'$, $\beta_{e(h)}=m_{e(h)}/(m_e+m_h)$. Similarly, $V_{n}^{\mrm{XX}}(Q,Q',q)$ reads
\begin{align}\label{intxx}
   & V_{n}^{\mrm{XX}}(Q,Q',q)=V_{n}^{\mrm{XX}}(q,\Delta Q=Q'-Q)= \\ \notag & \frac{1}{2L}\int \diff \tilde x \diff \tilde x' \diff \eta \, \mrm{e}^{i (-q+\Delta Q) \eta}|{\phi_{n}(\tilde x)}|^2|{\phi_{n}(\tilde x')}|^2 V_{\mrm{int}}(|\eta+\beta_h (\tilde x-\tilde x')|,|\eta-\beta_e (\tilde x-\tilde x')|,|\eta+\beta_h \tilde x +\beta_e \tilde x'|,|\eta-\beta_h \tilde x -\beta_e \tilde x'|),
\end{align}
where $\Delta Q$ is a difference between the initial momenta of two excitons.
The e-e exchange part is expressed as:
\begin{align}\label{intee}
   & V_{n}^{\mrm{ee}}(Q,Q',q)=V_{n}^{\mrm{ee}}(q,\Delta Q)= \\ \notag & \frac{1}{2L}\int \diff \tilde x \diff \tilde x' \diff \eta \, \mrm{e}^{i \beta_e \Delta Q (\eta+\beta_h(\tilde x-\tilde x'))}\mrm{e}^{i  q ((\beta_e-\beta_h)\eta-2\beta_h\beta_e(\tilde x-\tilde x'))}{\phi_{n}(|\eta+\beta_h \tilde x +\beta_e \tilde x'|)}{\phi_{n}(|\eta-\beta_h \tilde x -\beta_e \tilde x'|)}{\phi_{n}(\tilde x)}{\phi_{n}(\tilde x')} \\ \notag & \times  V_{\mrm{int}}(|\eta+\beta_h (\tilde x-\tilde x')|,|\eta-\beta_e (\tilde x-\tilde x')|,|\eta+\beta_h \tilde x +\beta_e \tilde x'|,|\eta-\beta_h \tilde x -\beta_e \tilde x'|).
\end{align}
Finally, the h-h exchange part reads:
\begin{align}\label{inthh}
   & V_{n}^{\mrm{hh}}(Q,Q',q)=V_{n}^{\mrm{hh}}(q,\Delta Q)= \\ \notag & \frac{1}{2L}\int \diff \tilde x \diff \tilde x' \diff \eta \, \mrm{e}^{i \beta_e \Delta Q (\eta-\beta_e(\tilde x-\tilde x'))}\mrm{e}^{-i  q ((\beta_e-\beta_h)\eta-2\beta_h\beta_e(\tilde x-\tilde x'))}{\phi_{n}(|\eta+\beta_h \tilde x +\beta_e \tilde x'|)}{\phi_{n}(|\eta-\beta_h \tilde x -\beta_e \tilde x'|)}{\phi_{n}(\tilde x)}{\phi_{n}(\tilde x')} \\ \notag & \times  V_{\mrm{int}}(|\eta+\beta_h (\tilde x-\tilde x')|,|\eta-\beta_e (\tilde x-\tilde x')|,|\eta+\beta_h \tilde x +\beta_e \tilde x'|,|\eta-\beta_h \tilde x -\beta_e \tilde x'|).
\end{align}
In this work, we consider the elastic scattering of two excitons whose initial momenta are small and comparable. Therefore, we assume that $\Delta Q \approx 0$. As an immediate consequence, we notice that
\begin{align}
   & V_{n}^{\mrm{dir}}(q)=V_{n}^{\mrm{XX}}(q,0), \\ 
   & V_{n}^{\mrm{ee}}(q,0)=V_{n}^{\mrm{hh}}(q,0).
\end{align}
Finally, the scattering amplitude $V_{n}(q,Q,Q')$ takes the form
\begin{equation}
    V_{n}(q,Q,Q')\approx V_{n}(q) = 2V_{n}^{\mrm{dir}}(q)+2V_{n}^{\mrm{ee}}(q).
\end{equation}
In the main text, we argue that in a typical experiment in TMDs excitons are created almost exclusively in the internal ground state. Then, we set $n=GS$ and define
\begin{equation}
V_{\mrm{eff}}(q) = V_{\mrm{dir}}(q)+V_{\mrm{exch}}(q)=2V_{GS}^{\mrm{dir}}(q)+2V_{GS}^{\mrm{ee}}(q).
\end{equation}
In the main text, we present our results for a concrete example of a $\mrm{WSe}_2$ monolayer, for which $m_e\approx m_h$ and thus $\beta_h=\beta_e=1/2$. Note, that our results also hold for $m_e \neq m_h$. The only difference is that integrals in Eq.~\eqref{intdirect}-\eqref{inthh} are more difficult to compute numerically. For their concrete computation, we set $V_{\mrm{int}}$ as
\begin{equation}
    V_{\mrm{int}}=
V_{\mrm{K}}\bb{x_{e}-x_{e'}}+V_{\mrm{K}}\bb{x_{h}-x_{h'}}-V_{\mrm{K}}\bb{\sqrt{\bb{x_{e}-x_{h'}}^2}+d^2}-V_{\mrm{K}}\bb{\sqrt{\bb{x_{e'}-x_{h}}^2}+d^2}.
\end{equation}
Lastly, one can simplify the integral given by Eq.~\eqref{intdirect} to a one-dimensional integral as shown in Ref.~\cite{shahnazaryan2017exciton}. In order to solve  the integral given by Eq.~\eqref{intee} numerically, we use a 
CUBA package~\cite{hahn2005cuba}.

\section{II. Analytical expressions for $\mathbf{\Delta E_{\mrm{GS}}(N)}$}

\subsubsection{(a) Tonks-Girardeau regime}
As pointed out by Girardeau in 1960~\cite{girardeau1960}, in the TG regime ($nr_0 \rightarrow 0$ for a system of dipoles), one can map interacting bosons onto a system of ideal fermions.
The ground state energy of $N$ noninteracting fermions on a periodic ring of circumference $L$ reads
\begin{equation}
    E_\mathrm{GS}^\mathrm{TG} =  \sum_{n=1}^{(N-1)/2} 2 \frac{(2 n\pi\hbar)^2}{2mL^2} =  \frac{\pi^2 \hbar^2}{2mL^2} \frac{N(N^2-1)}{6},
\end{equation}
%
where the scaling with $N$ originates from the summation over all contributing single-particle states, cf. Fig. \ref{Fig:FigureSM}(a).
Accordingly, the difference of GS energies between systems of $N$ and $N-1$ particles reads
\begin{equation}
   \Delta E_{\mrm{GS}}^{\mrm{TG}}(N)= E_{\mrm{GS}}^{\mrm{TG}}(N)-E_{\mrm{GS}}^{\mrm{TG}}(N-1)\sim\frac{\pi^2 \hbar^2}{mL^2} \frac{N^2}{2},
\end{equation}
which governs the scaling of the PL response in the TG regime as discussed in the main text.

\subsubsection{(b) Charge-density-wave regime}
In order to obtain the analytical scaling of the dominant PL response in the CDW regime we consider $N$ bosonic dipoles interacting via the two-body interaction potential 
$V_{\mrm{dip}}(r)=e^2 d^2/(4 \pi \varepsilon_0\epsilon r^3)$.
In units used throughout the main text, this potential takes the form $V(x) = r_0/(2x^3)$.
Approximating the system as dipoles being localized in their classical equilibrium configuration, the total energy has several contributions: as illustrated in Fig.~\ref{Fig:FigureSM}(b), for a system of $N$ dipoles there are $N$ potential energy contributions from nearest neighbours at distance $x=d$ ($d=L/N$), $N$ contributions from next-nearest neighbours at distance $x=2d$ etc.
Denoting the number of $k$-nearest neighbours by $m_k$, for sufficiently large $N$, the ground state energy then takes the form
%
\begin{figure}[t!]
\begin{center}
		\includegraphics[width=\textwidth]{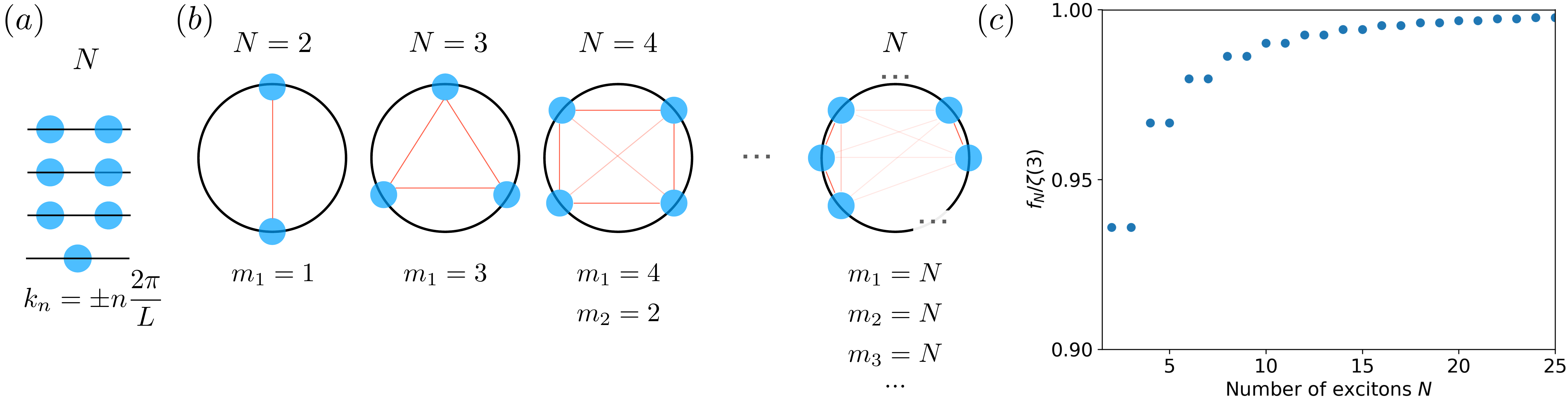}
\par\end{center}
\caption{Analysis of energy scalings $\Delta E_{\mrm{GS}}(N)$ in the TG and CDW limits.
(a) Occupied momenta $k_n = \pm n \pi / L$ ($n = 0, ..., (N-1)/2$) in a system of $N$ noninteracting fermions and their absolute values $|k_n|$.
(b) Number of $k$-nearest neighbours $m_k$ as system size $N$ grows.
The interaction of dipoles at a distance $x=kd$ is proportional to $1/(kd)^3$, indicated by brighter lines connecting dipoles that are farther away. 
(c) The ratio $f_N/\zeta(3)$ approaches unity rapidly for increasing values of the system size $N$.}
\label{Fig:FigureSM}
\end{figure}
\begin{equation}
    E_{\mrm{GS}}^{\mrm{CDW}}
    =\frac{r_0}{2}\sum_{k=1}^{\lfloor{}\frac{N-1}{2}\rfloor{}}\frac{m_k}{\bb{kd}^3}
    \approx \frac{r_0}{2d^3}\sum_{k=1}^{\lfloor{}\frac{N-1}{2}\rfloor{}}\frac{N}{k^3}
    =\frac{r_0 N^4}{2L^3} f_N,
\end{equation}
where in the last step we used the fact that $d=L/N$ and introduced $f_N = \sum_{k=1}^{\lfloor{}(N-1)/2\rfloor{}} 1/k^3$.
For large $N$, $f_N \rightarrow \zeta(3)$, as can be seen in Fig.~\ref{Fig:FigureSM}(c).
Thus, we finally obtain that the difference of GS energies between systems of $N$ and $N-1$ particles can be expressed as
\begin{equation}
   \Delta E_{\mrm{GS}}^{\mrm{CDW}}(N)= E_{\mrm{GS}}^{\mrm{CDW}}(N)-E_{\mrm{GS}}^{\mrm{CDW}}(N-1)\sim \frac{2r_0 \zeta(3)}{L^3} N^3,
\end{equation}
which is the scaling governing the PL response in the CDW regime discussed in the main text.

\section{III. Numerical details}

We access the many-body eigenstates of Hamiltonian~\eqref{eq:hamgen} by exact diagonalization using the Lanczos algorithm~\cite{lanczos1950iteration}. Our calculations are performed in the Fock space spanned by the plane-wave basis. We impose a cutoff on  the  maximum total kinetic  energy  of the system $E_{\rm max}=k_{\rm max}^2/2$ that corresponds to a maximum single-particle momentum $k_{\rm max}\gg 1/\sigma_\mathrm{X}$ sufficiently high to ensure convergence. Additionally, we make use of the fact that the total momentum of the system $\hat{K}=\sum_k \,k\, \hat{a}_{k}^\dagger \hat{a}_{k}$ is conserved, $[ \hat{H}, \hat{K} ]=0$. Thus, its eigenvalues $K$ are  good quantum numbers which we use ---together with the total number of atoms $N$--- to label the different eigenstates $\ket{N,\, K,\,i}$ of the system, enumerated by $i$. Specifically,  we focus on $K=0$ states, for which $i=0$ corresponds to the ground state.

\end{widetext}

\end{document}